# Fractional Variational Iteration Method for Fractional Nonlinear Differential Equations

Guo-cheng Wu*

Interdisciplinary Institute for Nonlinear Science, National Engineering Laboratory of Modern Silk, Soochow University, Suzhou, 215021

Abstract

Recently, fractional differential equations have been investigated via the famous variational iteration method. However, all the previous works avoid the term of fractional derivative and handle them as a restricted variation. In order to overcome such shortcomings, a fractional variational iteration method is proposed. The Lagrange multipliers can be identified explicitly based on fractional variational theory.

1  Introduction

The variational iteration method [1 – 4] has been extensively worked out for many years by numerous authors. Starting from the pioneer ideas of the Inokuti–Sekine–Mura method, Ji-Huan He [3] developed the variational iteration method (VIM) in (1999). In this method, the equations are initially approximated with possible unknowns. A correction functional is established by the general Lagrange multiplier which can be identified optimally via the variational theory. Besides, the VIM has no restrictions or unrealistic assumptions such as linearization or small parameters that are used in the nonlinear operators [5 - 11].

In the last three decades, scientists and applied mathematicians have found fractional differential equations useful in various fields: rheology, quantitative biology, electrochemistry, scattering theory, diffusion, transport theory, probability potential theory and elasticity [12]. Finding accurate and efficient methods for solving FDEs has been an active research undertaking. A question may naturally arise: Can we have a fractional variational method to derive approximate solutions of FDEs? Although a number of useful attempts have been made to solve fractional equations via the VIM, the problem has not yet been completely resolved, i.e., most of the previous works avoid the term of fractional derivative, handle them as restricted variation and they cannot identify the fractional Langrange multipliers explicitly in the correction function.

*Corresponding author, Email addresses: wuguocheng2002@yahoo.com.cn.

Assume the following fractional differential equation

$$\frac{D^\alpha u}{Dt^\alpha} + f(u,t) = 0, \quad a \leq t \leq b. \tag{1}$$

where $\frac{D^\alpha u(t)}{Dt^\alpha}$ is the famous Caputo's fractional derivative defined as

$$\frac{D^\alpha u(t)}{Dt^\alpha} = \frac{1}{\Gamma(m+1-\alpha)} \int_a^t \frac{u^{(m+1)}(\tau)}{(t-\tau)^{\alpha-m}} d\tau, \quad m < \alpha \leq m+1. \tag{2}$$

Here $\alpha$ is a real constant called the order of the fractional derivative, and $a$ is the initial point and $\Gamma$ denotes the gamma function.

In the case $0 < \alpha \leq 1$, we re-write Eq. (1) in the form

$$\frac{du}{dt} + \frac{D^\alpha u}{Dt^\alpha} - \frac{du}{dt} + f(u,t) = 0, \tag{3}$$

Handling the term $\frac{D^\alpha u}{Dt^\alpha} - \frac{du}{dt} + f(u,t)$ as restricted variation, we can find the following variational iteration algorithms [2]

$$\begin{cases} u_{n+1}(t) = u_n(t) - \int_0^t (\frac{D^\alpha u_n}{Dt^\alpha} + f_n) \, ds \\ u_{n+1}(t) = u_0(t) - \int_0^t (\frac{D^\alpha u_n}{Dt^\alpha} - \frac{du_n}{dt} + f_n) \, ds \\ u_{n+1}(t) = u_0(t) - \int_0^t \left\{ (\frac{D^\alpha u_n}{Dt^\alpha} - \frac{du_n}{dt} + f_n) - (\frac{D^\alpha u_{n-1}}{Dt^\alpha} - \frac{du_{n-1}}{dt} + f_{n-1}) \right\} ds. \end{cases} \quad (4)$$

In the case $1 < \alpha \leq 2$, the above iteration formulas are also valid. We can present Eq. (1) in the form

$$\frac{d^2 u}{dt^2} + \frac{D^\alpha u}{Dt^\alpha} - \frac{d^2 u}{dt^2} + f(u,t) = 0, \quad (5)$$

and the following iteration formulae are suggested [2]

$$\begin{cases} u_{n+1}(t) = u_n(t) + \int_0^t (s-t)(\frac{D^\alpha u_n}{Dt^\alpha} + f_n) \, ds \\ u_{n+1}(t) = u_0(t) + \int_0^t (s-t)(\frac{D^\alpha u_n}{Dt^\alpha} - \frac{d^2 u_n}{dt^2} + f_n) \, ds \\ u_{n+1}(t) = u_n(t) + \int_0^t (s-t) \left\{ (\frac{D^\alpha u_n}{Dt^\alpha} - \frac{d^2 u_n}{dt^2} + f_n) - (\frac{D^\alpha u_{n-1}}{Dt^\alpha} - \frac{d^2 u_{n-1}}{dt^2} + f_{n-1}) \right\} ds. \end{cases} \quad (6)$$

When $\alpha$ is close to 1, Eq. (4) is better while Eq. (6) is recommended for $\alpha$ approaching 2.

In the above derivation, the Lagrange multiplier in the correction functional is identified approximately using the classical variational principle, due to the lack of fractional variational theory. Now the things are changing. Recently, Jumarie proposed a modified Riemann-Liouville derivative [13]. With this kind of the fractional derivative, a generalized variational derivative is defined [13, 14], and variational approach for fractional partial differential equations is proposed [15]. Based on the fractional variational theory, we can extend the VIM to a fractional one and solve fractional differential equations with a generalized Lagrange multiplier.

2  Properties of Modified Riemann-Liouville Derivative

Comparing with the classical Caputo derivative, the definition of modified

Reimann-Liouville derivative is not required to satisfy higher integer-order derivative than $\alpha$. Secondly, $\alpha^{th}$ derivative of a constant is zero. Now we introduce some properties of the fractional derivative. Assume $f: R \to R$, $x \to f(x)$ denote a continuous (but not necessarily differentiable) function in the interval [0, 1]. Through the fractional famous Riemann Liouville integral

$$_0I_x^\alpha f(x) = \frac{1}{\Gamma(\alpha)} \int_0^x (x-\xi)^{\alpha-1} f(\xi) d\xi, \alpha > 0, \tag{7}$$

The modified Riemann-Liouville derivative is defined as [14]

$$_0D_x^\alpha f(x) = \frac{1}{\Gamma(1-\alpha)} \frac{d}{dx} \int_0^x (x-\xi)^{-\alpha} (f(\xi) - f(0)) d\xi, \tag{8}$$

where $x \in [0,1]$, $0 < \alpha < 1$.

In the next sections, we will use the following properties.

(a). Fractional Leibniz product law [13]

$$_0D_x^{(\alpha)}(uv) = u^{(\alpha)}v + uv^{(\alpha)}. \tag{9}$$

(b). Fractional integration by parts [13]

$$_aI_b^\alpha u^{(\alpha)} v = (uv)/_a^b - {_aI_b^\alpha} uv^{(\alpha)} \tag{10}$$

(c). Integration with respect to $(dx)^\alpha$ ( **Lemma 2.1** of [14])

We use the following equality for the integral w. r. t $(dx)^\alpha$

$$_0I_x^\alpha f(x) = \frac{1}{\Gamma(\alpha)} \int_0^x (x-\xi)^{\alpha-1} f(\xi) d\xi = \frac{1}{\Gamma(\alpha+1)} \int_0^x f(\xi)(d\xi)^\alpha, 0 < \alpha \le 1. \tag{11}$$

In order to propose a FVIM for fractional nonlinear equations, firstly, we should consider fractional variational theory which is employed to construct a fractional correction functional, then we need to identify the fractional Lagrange multipliers.

3  Fractional Variational Theory

Several versions of fractional variational approaches have been proposed [15 - 17]. However, all of them can not applied to establish a fractional variational functional for fractional differential equations. With Jumarie's fractional derivative [13 - 15], we can readily establish a generalized fractional functional. We now generally revisit the derivation of the fractional variational derivative [13, 14].  Start from the functional

$$J[y] = \frac{1}{\Gamma(1+\alpha)} \int_a^b F(x, y, {}_aD_x^\alpha y)(dx)^\alpha \qquad (12)$$

and find the necessary conditions for extrema. Let $y^*(x)$ be the desired function and let $\varepsilon \in R$. Let

$$y(x) = y^*(x) + \varepsilon \eta(x) \qquad (13)$$

be a family of curves that satisfy the boundary conditions, which we can set, for simplicity, as

$$\eta(a) = \eta(b) = 0.$$

As ${}_aD^\alpha y(x)$ is a linear operator, we have

$${}_aD^\alpha y(x) = {}_aD^\alpha y^*(x) + \varepsilon {}_aD^\alpha y^*(x)\eta(x), \qquad (14)$$

so that by substituting Eqs. (13) and (14) into Eq. (12), for each $\eta(x)$, we have

$$J = J[\varepsilon] = \frac{1}{\Gamma(1+\alpha)} \int_a^b F(x, y^* + \varepsilon\eta, {}_aD^\alpha y^* + \varepsilon {}_aD^\alpha \eta)(dx)^\alpha. \qquad (15)$$

Note that $J[\varepsilon]$ is a function of $\varepsilon$ only and it attains its extremum at $\varepsilon = 0$. Differentiating Eq. (15) with respect to $\varepsilon$ gives

$$\frac{dJ}{d\varepsilon} = \frac{1}{\Gamma(1+\alpha)} \int_a^b [\frac{\partial F}{\partial y}\eta + \frac{\partial F}{\partial {}_aD_x^\alpha y} {}_aD_x^\alpha \eta](dx)^\alpha, \qquad (16)$$

so that a necessary condition for $J(\varepsilon)$ to have an extremum is for $dJ/d\varepsilon$ to vanish for all admissible $\eta(x)$ which leads to the result

$$\frac{1}{\Gamma(1+\alpha)} \int_a^b [\frac{\partial F}{\partial y}\eta(x) + \frac{\partial F}{\partial {}_aD_x^\alpha y} {}_aD_x^\alpha \eta(x)](dx)^\alpha = 0. \qquad (17)$$

The integral in Eq. (17) can be rewritten, using the fractional Leibniz formula and integration by parts, in the form

$$\frac{1}{\Gamma(1+\alpha)}\int_a^b \frac{\partial F}{\partial_a D_x^\alpha \eta(x)} {}_a D_x^\alpha \eta(x)(dx)^\alpha$$
$$= \eta(x)\frac{\partial F}{\partial_a D_x^\alpha y(x)}\Big|_a^b - \frac{1}{\Gamma(1+\alpha)}\int_a^b \eta(x) {}_a D_x^\alpha \left(\frac{\partial F}{\partial_a D_x^\alpha y(x)}\right)(dx)^\alpha \quad (18)$$
$$= -\frac{1}{\Gamma(1+\alpha)}\int_a^b \eta(x) {}_a D_x^\alpha \left(\frac{\partial F}{\partial_a D_x^\alpha y(x)}\right)(dx)^\alpha,$$

so that we have

$$\frac{1}{\Gamma(1+\alpha)}\int_a^b \left[\frac{\partial F}{\partial y} - {}_a D_x^\alpha \left(\frac{\partial F}{\partial_a D_x^\alpha y(x)}\right)\right]\eta(x)(dx)^\alpha = 0. \quad (19)$$

As $\eta(x)$ is arbitrary, the Euler-Lagrange equation for the fractional variational principle is

$$\frac{\partial F}{\partial y} - {}_a D_x^\alpha \left(\frac{\partial F}{\partial_a D_x^\alpha y}\right) = 0. \quad (20)$$

Similarly, we can derive higher order fractional Euler-Lagrange equation

$$\frac{\partial F}{\partial y} + (-1)^k {}_a D_x^{k\alpha}\left(\frac{\partial F}{\partial_a D_x^{k\alpha} y}\right) = 0. \quad (21)$$

When $\alpha = 1$, Eq. (21) can turn out to be the Euler-Lagrange equation in usual sense.

4  Fractional Variational Iteration Method

In Ref. [18], we proposed a fractional VIM for two kinds of fractional diffusion equation. In this section, we solve two fractional nonlinear equations to illustrate the fractional iteration method's efficiency.

Example.1. As the first example, we consider a time-fractional diffusion equation. Previously, Oldham and Spanier [19] solved a fractional diffusion equation that contains first order derivative in space and half order derivative in time. Recently, F. Mainardi [20, 21] investigated analytically the time-fractional diffusion wave equations. In space fractional diffusion process, G. Gorenflo and F. Mainardi [22] obtained this fractional model by replacing the second order space derivative with a suitable fractional derivative operator.

The analytical fractional diffusion equation in time is governed by the equation [23]

$$\frac{\partial^{\alpha} u(x,t)}{\partial t^{\alpha}} = D\frac{\partial^2 u(x,t)}{\partial x^2} - \frac{\partial(F(x)u(x,t))}{\partial x}, \quad 0 < \alpha \le 1, \tag{22}$$

where $\frac{\partial^{\alpha}}{\partial t^{\alpha}}$ is the Caputo derivative, with initial condition $u(x,0) = f(x)$.

We replace the fractional Caputo derivative with the modified Riemann-Liouville derivative in Eq. (22), and assume $D = 1$, $F(x) = -x$ which leads to

$$\frac{\partial^{\alpha} u(x,t)}{\partial t^{\alpha}} = \frac{\partial^2 u(x,t)}{\partial x^2} + \frac{\partial(xu(x,t))}{\partial x}, \quad 0 < \alpha \le 1, \tag{23}$$

with initial condition $u(x,0) = x^2$.

Then a corrected functional for Eq. (23) can be constructed as follows

$$u_{n+1}(x,t) = u_n(x,t) + \frac{1}{\Gamma(1+\alpha)}\int_0^t \lambda(t,\tau)\{\frac{\partial^{\alpha} u_n(x,\tau)}{\partial \tau^{\alpha}} - \frac{\partial^2 \tilde{u}_n(x,t)}{\partial x^2} - \frac{\partial(x\tilde{u}_n(x,t))}{\partial x}\}(d\tau)^{\alpha}. \tag{24}$$

with the property from Eqs. (9 - 11), $\lambda(t,\tau)$ must satisfy

$$\frac{\partial^{\alpha} \lambda(t,\tau)}{\partial \tau^{\alpha}} = 0, \text{ and } 1 + \lambda(t,\tau)\big|_{\tau=t} = 0. \tag{25}$$

Therefore, $\lambda(t,\tau)$ can be identified as $\lambda(t,\tau) = -1$.

Substituting the initial value $u_0(x,t) = u_0(x,0) = f(x) = x^2$ into the iteration formulation as follows

$$u_{n+1}(x,t) = u_n(x,t) - \frac{1}{\Gamma(1+\alpha)}\int_0^t \{\frac{\partial^{\alpha} u_n(x,\tau)}{\partial \tau^{\alpha}} - \frac{\partial^2 u_n(x,t)}{\partial x^2} - \frac{\partial(xu_n(x,t))}{\partial x}\}(d\tau)^{\alpha}.$$

We can derive

$$u_1(x,t) = x^2 - \frac{1}{\Gamma(1+\alpha)}\int_0^t \{\frac{\partial^{\alpha} u_0(x,\tau)}{\partial \tau^{\alpha}} - \frac{\partial^2 u_0(x,t)}{\partial x^2} - \frac{\partial(xu_0(x,t))}{\partial x}\}(d\tau)^{\alpha}$$
$$= x^2 + \frac{(2+3x^2)t^{\alpha}}{\Gamma(1+\alpha)}, \tag{26}$$

$$u_2(x,t) = x^2 + \frac{(2+3x^2)t^{\alpha}}{\Gamma(1+\alpha)} + \frac{(8+9x^2)t^{2\alpha}}{\Gamma(1+2\alpha)}, \tag{27}$$

$$u_3(x,t) = x^2 + \frac{(2+3x^2)t^\alpha}{\Gamma(1+\alpha)} + \frac{(8+9x^2)t^{2\alpha}}{\Gamma(1+2\alpha)} + \frac{(26+27x^2)t^{3\alpha}}{\Gamma(1+3\alpha)}, \tag{28}$$

．．．．．．．

Finally, the exact solution is

$$u(x,t) = \lim_{n\to\infty} u_n(x,t) = \lim_{n\to\infty} \sum_{i=0}^{n} \frac{k^i t^{i\alpha}}{\Gamma(1+i\alpha)} = E_\alpha(kt^\alpha), \tag{29}$$

where $k^i = x^2 + (1+x^2)(3^i - 1)$.

If we assume $\alpha = \frac{1}{2}$ in Eq. (29), we can derive $E_{\frac{1}{2}}(k\sqrt{t})$ is the exact solution of the fractional diffusion equation [23].

Example.2. In order to illustrate the fractional VIM for higher fractional-order equations, we only consider the simple initial value problem in [23]

$$y^{(2\alpha)} = y^2 + 1, \quad 0 < \alpha \le 1, \quad 0 \le x \le 1. \tag{30}$$

$y(0)=0$ and $y^{(\alpha)}(0)=1$. Through this paper, $y^{(m\alpha)}$ is defined by $\underbrace{{}_aD_x^\alpha \cdots {}_aD_x^\alpha}_{m} y$.

Construct the following functional

$$y_{n+1}(x,t) = y_n(x,t) + \frac{1}{\Gamma(1+\alpha)} \int_0^x \lambda \{y_n^{(2\alpha)} - \tilde{y}_n^{\,2}(\xi) - 1\}(d\xi)^\alpha$$

$$\delta y_{n+1} = \delta y_n + \frac{1}{\Gamma(1+\alpha)} \delta \int_0^t \lambda(y_n^{(2\alpha)} - y_n^{2}(\xi) - 1)(d\xi)^\alpha$$

$$= \delta y_n + \lambda \delta y_n^{(\alpha)} |_{\xi=x} - \lambda^{(\alpha)}(\tau) \delta y_n(\tau)|_{\xi=x} + \frac{1}{\Gamma(1+\alpha)} \int_0^x \lambda^{(2\alpha)} \delta y_n (d\xi)^\alpha$$

Similarly, set the coefficients of $\delta u_n(\tau)$ and $\delta y_n^{(\alpha)}$ to zero

$$1 - \lambda^{(\alpha)}|_{\xi=x} = 0, \quad \lambda^{(2\alpha)} = 0 \quad \text{and} \quad \lambda|_{\xi=x} = 0.$$

Then, the generalized Lagrange multipliers for Eq. (31) can be identified as

$$\lambda = \frac{(\xi - x)^\alpha}{\Gamma(1+\alpha)}. \tag{31}$$

By the fractional variational theory and similar manipulation, we can derive more generalized Lagrange multipliers

$$\lambda = (-1)^{(m)} \frac{(\xi-x)^{(m-1)\alpha}}{\Gamma(1+(m-1)\alpha)},$$

for higher fractional nonlinear ordinary differential equations

$$y^{(m\alpha)} = N(y), \ 0<\alpha \leq 1.$$

From Eq. (31), we can check when $\alpha = 1$, $\lambda = \xi - x$ is the multiplier for the Riccati equation

$$y^{(')} = y^2 + 1. \tag{32}$$

The iteration formulation for Eq. (13) can be rewritten as

$$y_{n+1}(x) = y_n(x) + \frac{1}{\Gamma(1+\alpha)} \int_0^x \frac{(\xi-x)^\alpha}{\Gamma(1+\alpha)} \{y_n^{(2\alpha)} - y_n^2 - 1\}(d\xi)^\alpha.$$

With the fractional Jumarie-Taylor series [13] and the initial value

$$y_0 = y(0) + \frac{y^{(\alpha)}(0)x^\alpha}{\Gamma(1+\alpha)} = \frac{x^\alpha}{\Gamma(1+\alpha)},$$

We can obtain

$$y_1(x) = y_0(x) + \frac{1}{\Gamma(1+\alpha)} \int_0^x \frac{(\xi-x)^\alpha}{\Gamma(1+\alpha)} \{y_0^{(2\alpha)} - y_0^2 - 1\}(d\xi)^\alpha = \frac{x^\alpha}{\Gamma(1+\alpha)} + \frac{x^{2\alpha}}{\Gamma(1+2\alpha)},$$

$$y_2(x) = y_1(x) + \frac{1}{\Gamma(1+\alpha)} \int_0^x \frac{(\xi-x)^\alpha}{\Gamma(1+\alpha)} \{y_1^{(2\alpha)} - y_1^2 - 1\}(d\xi)^\alpha$$

$$= \frac{x^\alpha}{\Gamma(1+\alpha)} + \frac{x^{2\alpha}}{\Gamma(1+2\alpha)} + \frac{\Gamma(1+2\alpha)x^{4\alpha}}{\Gamma^2(1+\alpha)\Gamma(1+4\alpha)} + \frac{2\Gamma(1+3\alpha)x^{5\alpha}}{\Gamma(1+\alpha)\Gamma(1+2\alpha)\Gamma(1+5\alpha)}$$

$$+ \frac{\Gamma(1+4\alpha)x^{6\alpha}}{\Gamma^2(1+\alpha)\Gamma(1+6\alpha)}.$$

$y_2(x)$ is the second approximate form of Eq. (30). We can compare the solutions with the exact solutions through Figure.1 when assuming $\alpha \to 1$.

We note that the same result can be obtained by the Fractional Decomposition method [25].

## 5  Conclusion

Variational Iteration Method has proven as an efficient tool to solve nonlinear differential equations of integer order. In this paper, fractional variational iteration method is proposed. Now fractional nonlinear differential equations with modified Riemann-Liouville derivative can be solved by this technique.

**List of figure**

Fig.1 The 2nd approximate solution vs exact solution when $\alpha = 1$. the discontinuous line (--) is the approximate solution when $\alpha = 0.9$ and the dotted line (…) when $\alpha = 0.99$. The continuous line is the exact solution when $\alpha = 1$.

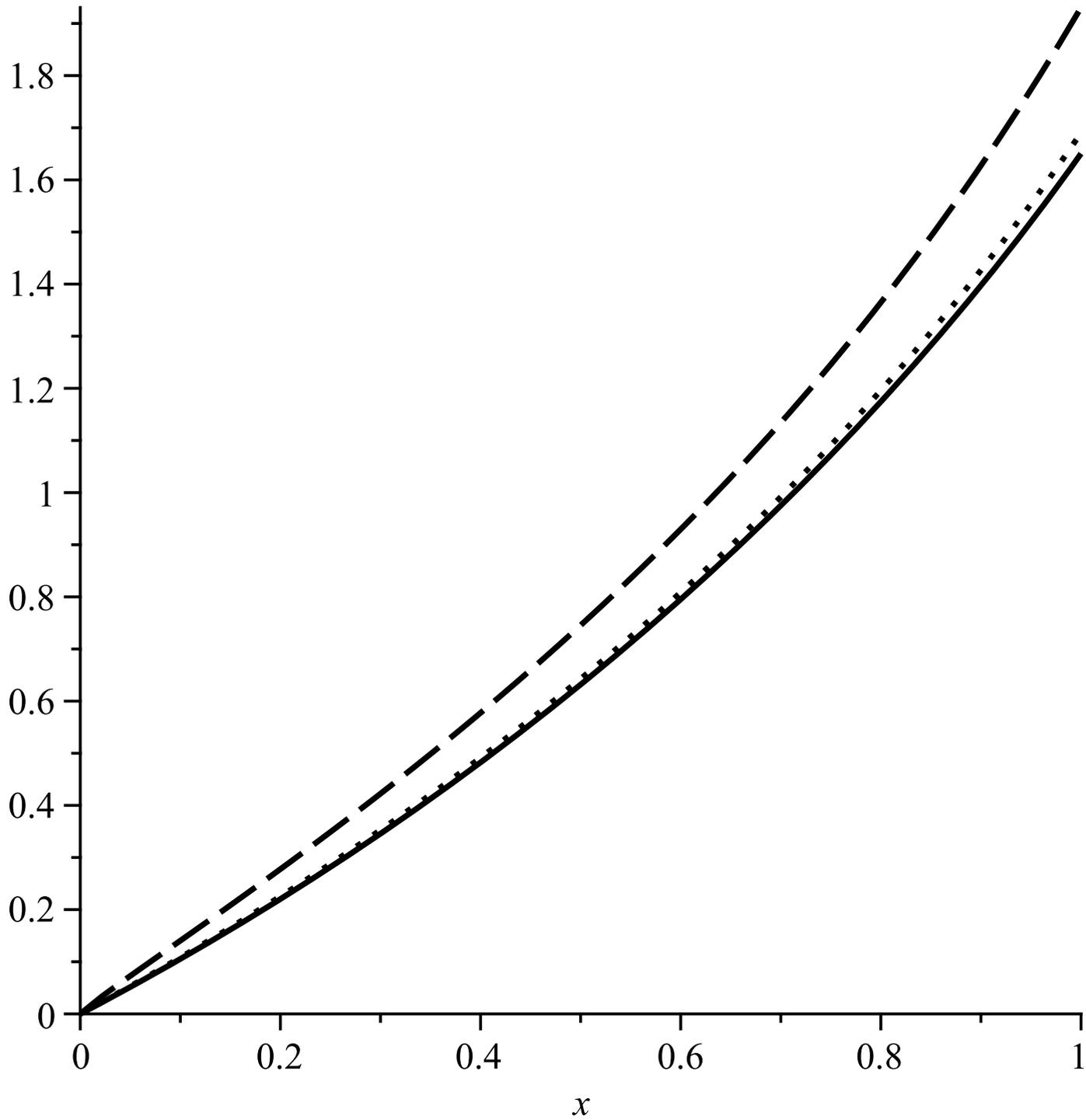